\documentclass[10pt, prl, aps, nofootinbib,
               superscriptaddress,
               amsmath, amssymb, twocolumn,
               preprintnumbers, showpacs,
               raggedbottom, floatfix]{revtex4-1}

\usepackage{bm}
\usepackage{graphicx}
\usepackage{xcolor}

\usepackage[colorlinks,breaklinks]{hyperref}
\DefineNamedColor{named}{RoyalBlue}{cmyk}{1,0.50,0,0}
\hypersetup{allcolors=RoyalBlue}

\newcommand{\beq}{\begin{gather}}
\newcommand{\eeq}{\end{gather}}
\newcommand{\op}[1]{#1}
\newcommand{\ket}[1]{\lvert #1\rangle}
\newcommand{\bra}[1]{\langle #1\rvert}
\newcommand{\braket}[1]{\langle #1\rangle}
\newcommand{\pdiff}[2]{\frac{\partial #1}{\partial #2}}

\let\Im\relax

\DeclareMathOperator{\Im}{Im}

\newcommand{\order}{O}
\newcommand{\vect}[1]{\boldsymbol{#1}}

\begin{document}

\title{Quantum Friction: Cooling Quantum Systems with Unitary Time Evolution}

\author{Aurel Bulgac}
\affiliation{Department of Physics, University of Washington, Seattle,
  Washington 98195--1560, USA}

\author{Michael McNeil Forbes}
\affiliation{Institute for Nuclear Theory, University of Washington,
  Seattle, Washington 98195--1550, USA}
\affiliation{Department of Physics, University of Washington, Seattle,
  Washington 98195--1560, USA}
\affiliation{Department of Physics \& Astronomy, Washington State University,
  Pullman, Washington 99164--2814, USA}

\author{Kenneth J. Roche}
\affiliation{Pacific Northwest National Laboratory,
  Richland, Washington 99352, USA}
\affiliation{Department of Physics, University of Washington, Seattle,
  Washington 98195--1560, USA}

\author{Gabriel Wlaz\l{}owski}
\affiliation{Faculty of Physics, Warsaw University of Technology,
  Ulica Koszykowa 75, 00-662 Warsaw, Poland}
\affiliation{Department of Physics, University of Washington, Seattle,
  Washington 98195--1560, USA}
  
\begin{abstract}\noindent
  We introduce a type of quantum dissipation -- local quantum friction -- by
  adding to the Hamiltonian a local potential that breaks time-reversal
  invariance so as to cool the system. Unlike the Kossakowski-Lindblad master
  equation, local quantum friction directly effects unitary evolution of the
  wavefunctions rather than the density matrix: it may thus be used to cool
  fermionic many-body systems with thousands of wavefunctions that must remain
  orthogonal.  In addition to providing an efficient way to simulate quantum
  dissipation and non-equilibrium dynamics, local quantum friction coupled with
  adiabatic state preparation significantly speeds up many-body simulations,
  making the solution of the time-dependent Schr\"odinger equation significantly
  simpler than the solution of its stationary counterpart.
\end{abstract}

\date{\today}

\preprint{INT-PUB-13-018}
\preprint{NT@UW-13-20}

\maketitle
\noindent
Quantum dissipation in strongly coupled systems -- where a subsystem
irreversibly loses energy to the surrounding environment -- has been studied for
over half a century, but progress beyond simple systems has been slow due to its
cumbersome theoretical framework. In addition to the formal difficulties of
deriving accurate descriptions for interesting strongly interacting systems that
lack a small expansion parameter -- nuclear and chemical reactions, transport
phenomena, response to very intense external fields, etc.\@ -- the evolution
equations for the reduced density matrix of the subsystem of interest have
strong memory effects~\cite{Feynman:1963, Baym:1961, *Kadanoff:1962,
  Keldysh:1964ud, *Keldysh:1964ud_en, Caldeira:1981}.  This makes even numerical
implementations of the derived master equations prohibitive in three dimensions
(3D) because of the memory required to store the density matrix with a long
history. In this letter, we propose a simple and efficient technique for quantum
dissipation via unitary evolution with a local quantum friction potential which
has no memory effects.  Unlike the Kossakowski-Lindblad (KL) master equation
\cite{Kossakowski:1972, *Lindblad:1976} the local quantum friction effects
strictly unitary evolution of wavefunctions.  We apply this to efficiently
generate initial configurations for density functional theories
(DFTs)~\cite{HK:1964, *Kohn:1965fk, *Kohn:1999fk, *Dreizler:1990lr,
  *Parr:1989uq}, thereby solving a long-standing problem in applying
time-dependent DFT (TDDFT)~\cite{Runge:1984mz, *Gross:2006, *Gross:2012} to
large scale systems ($N \sim 10^5$ to $10^6$ single particle states on a
three-dimensional (3D) lattice). Traditional methods require diagonalizing the
single-particle Hamiltonian which takes $\order(N^3)$ operations per iteration
with significant communication requirements on parallel computers.  In contrast,
real-time evolution scales as $\order(N^2\log N)$.  The use of local quantum
friction makes solving the time-dependent Schr\"odinger equation significantly
simpler than the solving its stationary counterpart.  Thus, the procedure we
describe may advance several fields of physics where fermionic systems can
be studied with TDDFT, including nuclear physics (nuclei and neutron stars),
trapped cold fermionic atoms, and electronic structure.

Modern approaches to the nuclear structure of medium to heavy nuclei rely on
density functional theory (DFT): an in principle exact approach~\cite{HK:1964,
  *Kohn:1965fk, *Kohn:1999fk, *Dreizler:1990lr, *Parr:1989uq} that includes and
extends mean-field techniques. DFT and its time-dependent extension TDDFT, see
e.g.~\cite{Runge:1984mz, *Gross:2006, *Gross:2012, Yabana:1996, Pohl:2000,
  Fennel:2010}, provide a unified approach to study both structure and reactions
(dynamics).  With more than two decades of success describing normal electronic
systems, TDDFT and the formally related time-dependent Hartree-Fock approach
have been used extensively in nuclear physics (see, for example,
\cite{Flocard:1978, *Negele:1978, *Simenel:2011, *Oberacker:2010,
  *Broomfield:2008, *Nakatsukasa:2012}).  Including pairing correlations extends
the application to cold atomic gases (via the superfluid local density
approximation (SLDA)~\cite{Bulgac:2011, Bulgac:2013b, Bulgac:2011b,
  Bulgac:2011c, Bulgac:2009}) and additional nuclear systems (via the
Hartree-Fock-Bogoliubov (HFB) and Skyrme functionals~\cite{Bulgac:2011,
  Bulgac:2013b, Bulgac:2011b, Bulgac:2011c, Bulgac:2009, Stetcu:2011}): Pairing
correlations are essential to describe low-energy induced fission of medium and
heavy nuclei, for example.

The implementation of TDDFT in 3D without symmetry restrictions has remained an
outstanding problem due to the numerical complexity of the problem. For example,
consider the fission of a heavy nucleus into two spatially separated fragments:
at a minimum, one requires a volume of $40 \times 40 \times 60$~fm$^3$ with a
resolution of at least $1$~fm (corresponding to an energy cutoff of $\sim 200$
MeV).  A single particle wavefunction thus requires $N\sim 400\,000$ components
including pairing and spin degrees of freedom.  Correctly describing
time-dependent pairing correlations requires essentially the full
single-particle spectrum, so the construction of an initial state with
conventional iterative techniques requires repeatedly diagonalizing the $N\times
N$ single-particle Hamiltonian (an $\order(N^3)$ operation) for the hundreds of
iterations required to converge to the self-consistent ground state with
sufficient accuracy for starting numerically stable TDDFT time-integration
algorithms~\cite{Bulgac:2011, Bulgac:2013b, Bulgac:2011b, Bulgac:2011c,
  Bulgac:2009, Stetcu:2011, takashi}.  Nonlinear optimization methods help, see
Broyden's method \cite{Baran;Bulgac;Forbes;Hagen;Nazarewicz...:2008-05}, but
still require hundreds of iterations, and convergence can remain elusive when
the $N \gtrsim 10^5$.

To put this in perspective, a single diagonalization for both proton and neutron
single-particle Hamiltonians took essentially the entire (now retired) JaguarPF
computer -- $217\,800$ of the $224\,256$ processor cores -- about 6 hours of
wall-time (about one million CPU hours), or about a month to determine just the
initial state.  Extending this approach to study nuclear matter in the crust of
neutron stars requires approximately an order of magnitude more states, and
hence will take hundreds to thousands of times longer which is prohibitively
expensive, even with exascale computing.  In contrast, the time-dependent
problem is significantly less demanding: Lattice based approaches can leverage
the Fast Fourier Transform (FFT) to achieve $\order(N^2\ln N)$ operations per
iteration.  For the SLDA, a problem of similar magnitude can be evolved for
around $10^6$ timesteps in a single day.  Leveraging general purpose graphics
processing units (GPUs), the runtime can be reduced by an order of magnitude.

One can leverage efficient real-time evolution to prepare the ground state of
some Hamiltonian $H_1$ via adiabatic state preparation: Start with the ground
state of a solvable Hamiltonian $H_0$, then slowly evolve with the
time-dependent Hamiltonian $H_t = s_tH_1 + (1-s_t)H_0$ where $s_t$ is a smoothly
varying switching function that goes from $s_{t\leq 0} = 0$ to $s_{t\geq T} = 1$
over a period of time $T$.  In the adiabatic limit, the system will track the
ground state of $H_t$, ending up in the ground state of $H_1$ with controllable
accuracy.  While effective~\cite{Pfitzner:1994, Bulgac:2013a}, this method has
some difficulties: narrowly avoided level crossings and their associated
Landau-Zener transitions can require long evolution times, otherwise the
non-adiabatic evolution will introduce excess energy into the system and produce
an excited state.

We introduce the local quantum friction potential to solve this problem,
enabling shorter preparation periods $T$.  All quantities here are local and
the position dependence is suppressed to simplify the formulas.  A
time-dependent external potential $U_t \equiv U_t(\vect{x})$ can do work on a
system, but can also be constructed to remove energy from a ``moving'' (in the
sense of having currents) system. The wavefunction $\Psi_t$ evolves via the
Schr\"odinger equation,
\begin{gather}\label{eq:SEQ}
  i\hbar\dot{\Psi}_t \equiv i\hbar\pdiff{\Psi_t}{t} = (H_t + U_t)\Psi_t,
\end{gather}
from which one can deduce the change in the energy $E =
\bra{\Psi_t}H_t\ket{\Psi_t}$ to be $\dot{E} = \braket{\dot{H}_t} +
2\Im\bra{\Psi_t}H_tU_t\ket{\Psi_t}/\hbar$.  The local quantum friction potential
$U_t$ therefore cools the system $\dot{E} \leq \braket{\dot{H}_t}$, removing
energy imparted by the switching process, if
\begin{gather}\label{eq:U}
  U_t \propto -2\Im(\Psi_t^*H_t\Psi_t)
  = -\hbar\,\vect{\nabla}\cdot\vect{j}_t = \hbar\,\dot{\rho}_t,
\end{gather}
where $\vect{j}_t = \hbar\Im(\Psi_t^*\vect{\nabla}\Psi_t)/m$ is the probability
current density and assumes standard kinetic energy $K = -\hbar^2\nabla^2/2m$
and a local potential.  The last form follows immediately from the continuity
equation. A local friction potential $U_t$ of this type explicitly breaks
time-reversal invariance.  Intuitively, $U_t$ removes any irrotational currents
in the system, damping currents by being repulsive where they are converging.
This continuous cooling thus removes any excess energy introduced as currents by
the state preparation $H_t$ in a manner similar to classical friction -- the
faster the system moves, the more currents are generated, and the faster $U_t$
will cool the system.

The local density $\rho_t$ provides a natural way to normalize the
potential~\eqref{eq:U} that works well in practice:
\begin{gather}\label{eq:LQFP}
  U_{t} = -\beta \frac{\hbar\,\vect{\nabla}\cdot\vect{j}_t}{\rho_t} 
  = \beta \frac{\hbar\,\dot{\rho}_t}{\rho_t},
\end{gather}
where $\beta$ is a dimensionless constant of order unity. Thus,
we have an efficient and practical method for preparing arbitrary initial states
in quantum systems: Proceed as with adiabatic state preparation, but evolve with
the Hamiltonian $H_t + U_t$ including the local quantum friction
potential~\eqref{eq:LQFP}, which explicitly breaks time-reversal invariance, and
removes any irrotational currents generated by faster-than-adiabatic
evolution. After sufficient evolution -- generally much shorter than required by
pure adiabatic state preparation -- the system will cool to a state without
irrotational currents.  In principle, this final state may be some excited
eigenstate of the system, but evolving from the ground state of $H_0$ generally
ensures convergence to the ground state of $H_1$.

The beauty of local quantum friction is its implementation. $U_t$ can be applied
in exactly the same way as any local external potential contained in $\op{H}_t$,
and manifestly preserves unitary evolution; all evolved wavefunctions remain
normalized and orthogonal. It can thus be applied directly to fermionic TDDFTs
which must evolve hundreds of thousands of single-particle wave functions
$\psi_n(t)$ while preserving orthonormality $\braket{\psi_m(t)|\psi_n(t)}
= \delta_{mn}$.  Each wavefunction evolves with the same single-particle
Hamiltonian $H_t + U_t$, but the density $\rho_t$ and current density
$\vect{j}_t$ are now traces over the Slater determinant of states:
\begin{align}
  \vect{j}_t &= \frac{\hbar}{m} 
  \Im \sum_{n=1}^N \psi^*_n(t)\vect{\nabla}\psi_n(t), &
  \rho_t &= \sum_{n=1}^N\psi_n^*(t)\psi_n(t).
\end{align}
In this manner, a set of orthogonal single-particle wave functions constructed
for the trivial initial Hamiltonian $H_0$ is continuously transformed into
another set of orthogonal single-particle wave functions finally describing the
self-consistent ground state of the non-trivial Hamiltonian $H_1$.  In other
words, the evolution evolves one Slater determinant into another Slater
determinant.  This is not complicated by the fact that for TDDFT, the
single-particle Hamiltonian $H_t$ depends non-linearly on the single-particle
wavefunctions -- the evolution remains unitary.

Note that one need not start with eigenstates of $H_0$. In principle, \emph{any}
set of orthogonal wavefunctions can be used to start the process and local
quantum friction~\eqref{eq:SEQ} will eventually bring the system to a
self-consistent state devoid of irrotational currents.  In practice, starting
with a poor choice of initial wavefunctions may stall or terminate the process
near stationary excited states of $H_1$: The use of quasi-adiabatic state
preparation from the ground state of $H_0$ generally ensures a more rapid
cooling to the ground state of $H_1$.

\begin{figure}[tbp]
  \includegraphics[width=\columnwidth]{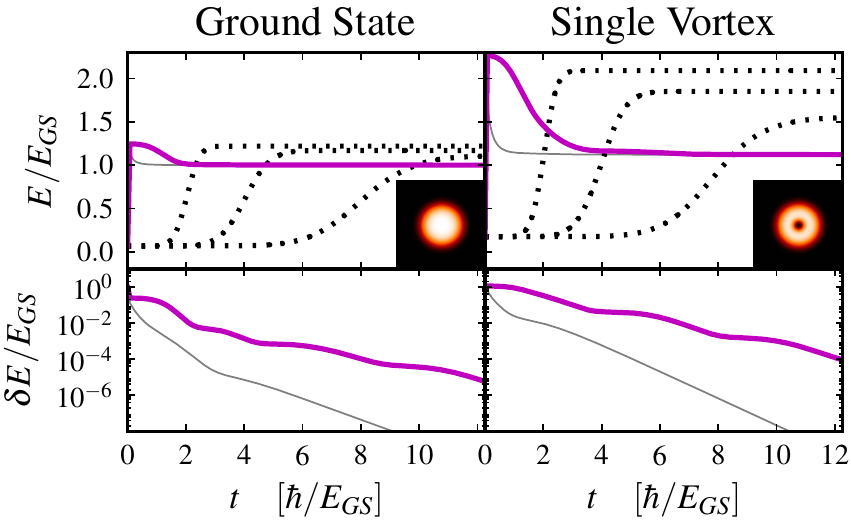}
  \caption{(Color online) \label{fig:GPE} Preparation of the ground state (GS)
    (left) and the single vortex state (right) with the Gross-Pitaevskii
    equation (GPE) (density profiles shown inset).  Dotted lines show adiabatic
    state preparation from exact HO eigenstates: slower preparations have lower
    final energy.  Solid lines show the local cooling algorithm from the HO
    initial state.  The thin grey line shows imaginary time evolution which is
    not practical for fermionic DFTs.  The lower subplots demonstrate the
    convergence on a logarithmic scale: like imaginary time evolution, quantum
    friction can exhibit near exponential convergence.
    }
\end{figure}

We shall now illustrate the power of this approach with three examples of
increasing numerical complexity. In the first example (Fig.~\ref{fig:GPE}), we
prepare the ground state of a Bose gas in a quartic trap with and without a
vortex.  The ground state and first Landau-level for a non-interacting gas in a
harmonic trap $H_0$ evolve into the ground state and single vortex state of the
interacting gas in a quartic trap $H_1$.  With only a single wavefunction, one
does not typically observe stalling near states with no irrotational currents,
so adiabatic state preparation is neither needed nor helpful -- instead the
system is cooled directly from the ground-state of $H_0$ after rapidly quenching
the system to $H_1$.  In this case only, one may compare with imaginary time
evolution finding that local quantum friction is somewhat slower, but still
exhibits the near exponential convergence.

In the second example, we determine the lowest 20 eigenstates of a symmetrized
Woods-Saxon potential in three dimensions.  We discretize the Schr\"odinger
equation using the discrete variable representation (DVR) technique described in
Ref.~\cite{Bulgac:2013} on a $32^3$ spatial lattice.  The Hamiltonian matrix has
$32\,768^2$ entries and is cumbersome to diagonalized on a laptop. Evolving with
local quantum friction, however, the 20 lowest bound states can be found in
under half an hour. We solve the problem with and without quasi-adiabatic
switching and, although direct cooling works, quasi-adiabatic switching improves
the efficiency allowing one to evolve for shorter periods $T$.

\begin{figure}[tbp]
  \includegraphics[width=\columnwidth]{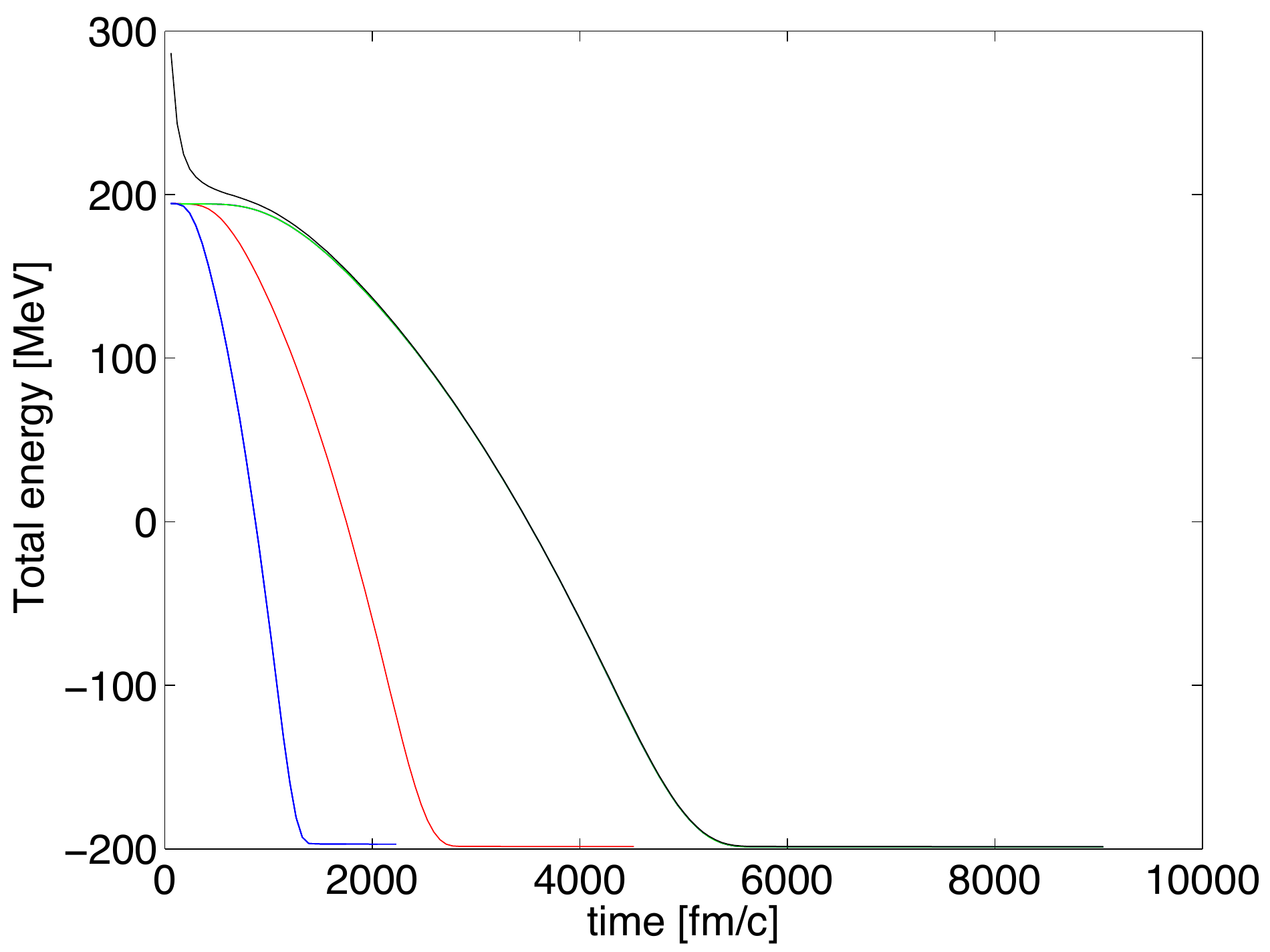}
  \caption{(Color online) \label{fig:WS}%
    The total instantaneous energy of a system of twenty non-interacting
    neutrons evolving from an initial 3D harmonic oscillator potential to a
    final symmetrized Woods-Saxon potential.  The curves correspond to
    quasi-adiabatic evolution with friction $(1-s_t)H_0 + s_t H_1 + U_t$ for
    various switching periods $T$ (two-thirds of the simulation time) and just
    friction $H_1+U_t$ for the remaining third of the simulation.  That the
    energy is constant during this time demonstrates that the ground state has
    been reached.  Note: there are three curves for the longest $T$
    corresponding to different simulations with $\{24^3, 32^3, 40^3\}$ lattices
    of 1~fm spacing: this demonstrates the infrared (IR) convergence.}
\end{figure}

In the final example, we demonstrate local quantum friction applied to a
large-scale TDDFT simulation modeling the unitary Fermi gas with the
TDSLDA~\cite{Bulgac:2011, Bulgac:2013b, Bulgac:2011b, Bulgac:2011c,
  Bulgac:2009}. We start with a self-consistent solution in an axially symmetric
trap $V_0(x,y) = m\omega_\perp^2(x^2+y^2)/2$ with periodic boundary conditions
along $z$. Note that translational invariance in $z$ renders this an effectively
two-dimensional problem that can be solved using traditional approaches.  Using
the combination of adiabatic switching and local quantum friction, we evolve to
the ground state in an elongated three-dimensional trap $V_1(x,y,z) =
m\omega_\perp^2(x^2+y^2)/2 + m\omega_\parallel^2z^2/2$ with a 1:4 aspect ratio
in a box with $16\times16\times64$ spatial lattice points.  The single-particle
Hamiltonian here is a $32\,768\times32\,768$ matrix, and $15\,322$ coupled
nonlinear partial differential equations (PDEs) in 3D and time are solved. The
solution -- using twenty four GPUs on the UW Hyak cluster -- takes about one
hour to converge to the ground state.  The final configuration of Titan
(NCCS)~\cite{titan} will have 18\,680 GPUs, and we expect to be able to solve
several million PDEs on spatial lattices of $\sim 100^3$ points.

\begin{figure}[tbp]
  \includegraphics[angle=-90,width=\columnwidth]{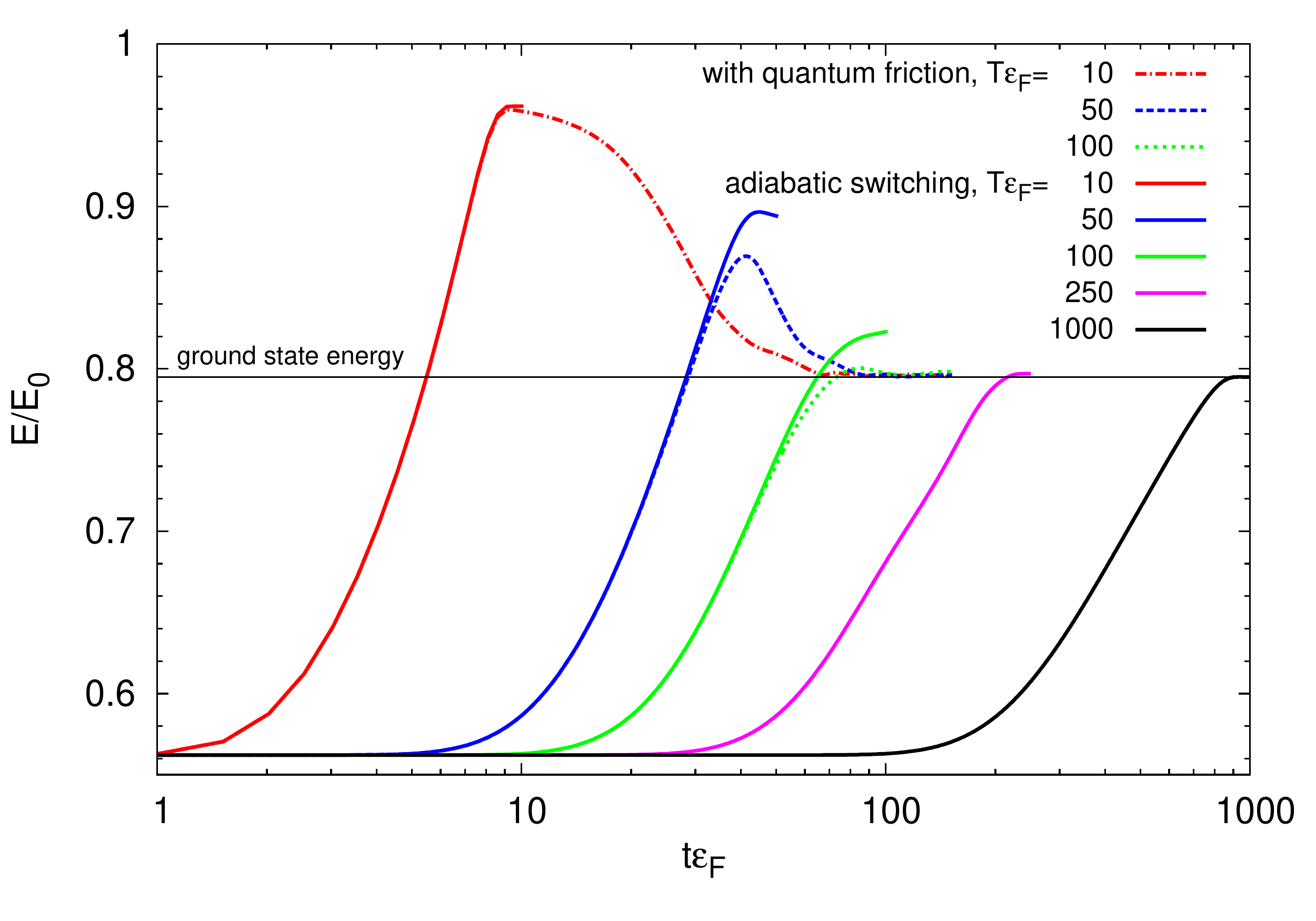}
  \caption{(Color online) \label{fig:UFG}%
    Ground state preparation of a trapped three-dimensional UFG gas from a
    translationally invariant (effectively two-dimensional) solution. The ground
    state can be prepared purely using adiabatic switching (solid lines) but the
    switching period can be dramatically reduced with local quantum
    friction. $E_{0}=\frac{3}{5}N\varepsilon_{F}$, where $N$ is number of
    particles and $\varepsilon_{F}$ is Fermi energy in the center of initial
    solution. (See also the movie in~\cite{EPAPS}.)}
\end{figure}

A natural extension to including local quantum friction is to include a
fluctuating potential to create a quantum equivalent of the Langevin equation.
By adjusting the nature and amplitudes of the fluctuating and dissipating terms,
one can generate different ensembles.  Note that this is similar to the
stochastic projected GPE (SPGPE), see~\cite{Rooney:2012} and references therein.
Efficient real-time evolution of this modified Schr\"odinger equation can then
be used as a tool to study non-equilibrium quantum systems at finite
temperatures. Local quantum friction might provide a more efficient approach to
study quantum dissipation than the Markovian Kossakowski-Lindblad master
equation, and provide an ideal tool for simulating non-equilibrium dynamics
where interactions with the environment are modeled by the time-dependent local
quantum friction potential $U_t$ and a stochastic driving term.  This
interaction is essentially classical, lacking entanglement between the quantum
system and the environment.  As such, the implementation is a simple extension
of the time-dependent Schr\"odinger equation for wavefunctions and one does not
have to evolve a reduced density matrix as in traditional approaches to quantum
dissipative evolution~\cite{Feynman:1963, Baym:1961, *Kadanoff:1962,
  Keldysh:1964ud, *Keldysh:1964ud_en, Caldeira:1981, Kossakowski:1972,
  *Lindblad:1976}.

We have shown how quantum friction can be introduced through a local potential
that may be efficiently computed using standard techniques.  The resulting
evolution is manifestly unitary, preserving the orthonormality of states. This
allows one to leverage efficient algorithms for simulating real-time dynamics
with time-dependent density functional theories (TDDFTs), and we have
demonstrated the efficacy of combining quasi-adiabatic state preparation with
local quantum friction to prepare initial states.  This algorithm greatly
extends the reach of quantum many-body simulations since solving the
time-dependent Schr\"odinger equation becomes easier than solving its static
counterpart.
  
\acknowledgments

\providecommand{\MMFGRANT}{\MakeUppercase{de-fg02-00er41132}}

We acknowledge support under U.S. Department of Energy (DoE) Grant
Nos. DE-FG02-97ER41014 and \MMFGRANT, and Contract No.\@ N\,N202\,128439 of the
Polish Ministry of Science. G.W.\@ acknowledges the Polish Ministry of Science
for the support within the program ``Mobility Plus \!-\! I edition'' under
Contract No.\@ 628/MOB/2011/0. Some of the calculations reported here have been
performed at the University of Washington Hyak cluster funded by the NSF MRI
Grant No.\@ PHY-0922770. This research also used resources of the National
Center for Computational Sciences at Oak Ridge National Laboratory, which is
supported by the Office of Science of the DoE under Contract
DE-AC05-00OR22725.


%
\end{document}